\documentclass[conference]{IEEEtran}
\IEEEoverridecommandlockouts
\usepackage{cite}
\usepackage{amsmath,amssymb,amsfonts}
\usepackage{algorithmic}
\usepackage{graphicx}
\usepackage{textcomp}
\usepackage{xcolor}
\def\BibTeX{{\rm B\kern-.05em{\sc i\kern-.025em b}\kern-.08em
    T\kern-.1667em\lower.7ex\hbox{E}\kern-.125emX}}
\begin{document}

\title{Reducing Uncertainty in Sea-level Rise Prediction: A Spatial-variability-aware Approach\\
}

\author{\IEEEauthorblockN{Subhankar Ghosh}
\IEEEauthorblockA{\textit{Dept. of Computer Science \& Eng.} \\
\textit{University of Minnesota, Twin Cities}\\
Minneapolis, USA \\
ghosh117@umn.edu}
\and
\IEEEauthorblockN{Shuai An}
\IEEEauthorblockA{\textit{Dept. of Computer Science \& Eng.} \\
\textit{University of Minnesota, Twin Cities}\\
Minneapolis, USA \\
an000033@umn.edu}
\and
\IEEEauthorblockN{Arun Sharma}
\IEEEauthorblockA{\textit{Dept. of Computer Science \& Eng.} \\
\textit{University of Minnesota, Twin Cities}\\
Minneapolis, USA \\
sharm485@umn.edu}
\and
\IEEEauthorblockN{Jayant Gupta}
\IEEEauthorblockA{\textit{Dept. of Computer Science \& Eng.} \\
\textit{University of Minnesota, Twin Cities}\\
Minneapolis, USA \\
gupta423@umn.edu}
\and
\IEEEauthorblockN{Shashi Shekhar}
\IEEEauthorblockA{\textit{Dept. of Computer Science \& Eng.} \\
\textit{University of Minnesota, Twin Cities}\\
Minneapolis, USA \\
shekhar@umn.edu}
\and
\IEEEauthorblockN{Aneesh Subramanian}
\IEEEauthorblockA{\textit{Dept. of Atmospheric \& Oceanic Sciences} \\
\textit{University of Colorado, Boulder}\\
Boulder, USA \\
aneeshcs@colorado.edu}
}

\maketitle

\begin{abstract}
Given multi-model ensemble climate projections, the goal is to accurately and reliably predict future sea-level rise while lowering the uncertainty. This problem is important because sea-level rise affects millions of people in coastal communities and beyond due to climate change's impacts on polar ice sheets and the ocean. This problem is challenging due to spatial variability and unknowns such as possible tipping points (e.g., collapse of Greenland or West Antarctic ice-shelf), climate feedback loops (e.g., clouds, permafrost thawing), future policy decisions, and human actions. Most existing climate modeling approaches use the same set of weights globally, during either regression or deep learning to combine different climate projections. Such approaches are inadequate when different regions require different weighting schemes for accurate and reliable sea-level rise predictions. This paper proposes a zonal regression model which addresses spatial variability and model inter-dependency. Experimental results show more reliable predictions using the weights learned via this approach on a regional scale.
\end{abstract}


\section{Introduction}
Sea-level rise is one of the most pressing environmental challenges facing the world today. In 2021 Intergovernmental Panel on Climate Change (IPCC) \cite{b26} projected that the global sea level is likely to rise by 10 to 32 inches (26 - 82 cm) by 2100, relative to 1986-2005. However, these global projections mask significant regional variation: sea level is projected to rise more in the Arctic than in the tropics and more in the mid-latitudes than in the subtropics. For example, in Chesapeake Bay (near Maryland and Washington DC), projections based on tide gauges, satellite observations and geo-physical models indicate that sea level will probably rise between 12-20 inches (30 - 50 cm) by 2050 (starting from 2005) \cite{b25}. The need for regional sea-level rise projections is becoming increasingly urgent. Coastal communities worldwide are already experiencing the impacts of sea-level rise, such as flooding, erosion, and saltwater intrusion. These impacts are expected to worsen as the sea level continues to rise. Regional sea-level rise projections are essential for coastal planning and adaptation. They can help communities identify areas that are most at risk and develop strategies to mitigate and adapt to these impacts.

Given a multi-model ensemble of climate projections from state-of-the-art climate models of the present, the goal is to accurately and reliably predict future sea-level rise while lowering the prediction uncertainty. A spatially-variable weighting scheme assigns different sets of weights to the candidate models in different regions unlike global climate models (GCMs), which assign the same set of weights across different regions. Using a spatially variable weighting scheme ensures that higher priority is given to more reliable models in a specific region when performing predictions and reducing uncertainty.

\begin{figure*}[ht]
    \centering
    \includegraphics[width=0.9\textwidth]{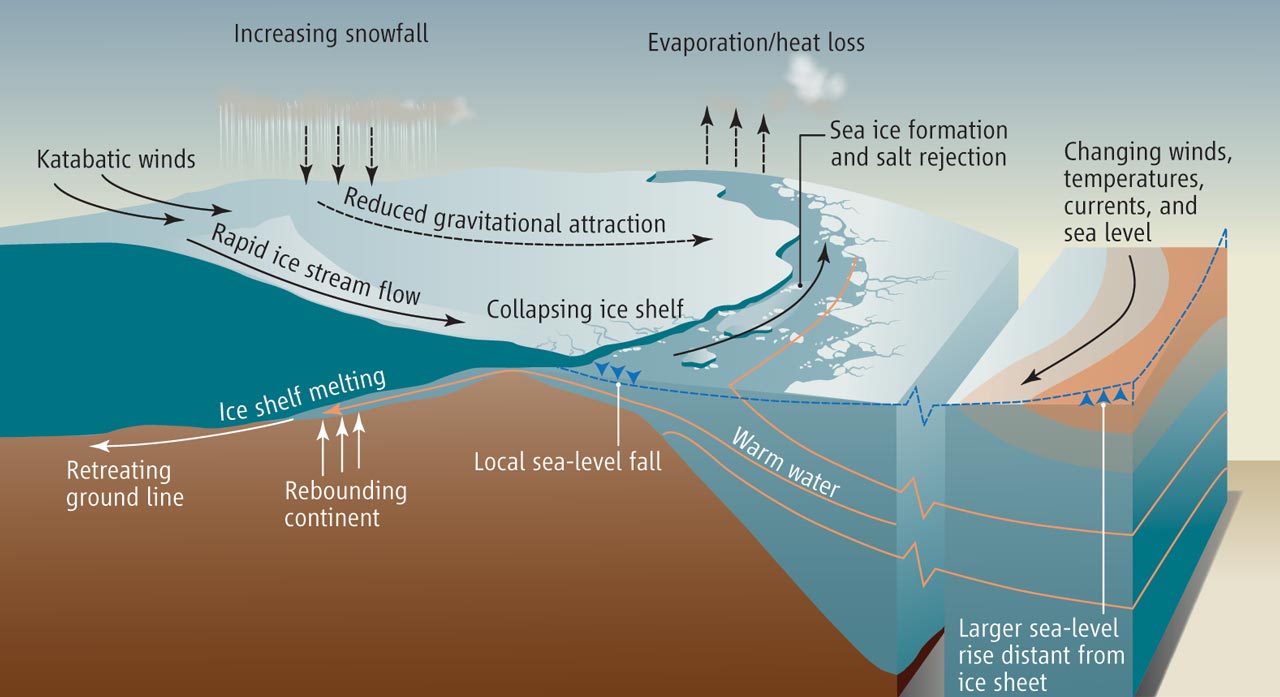}
    \caption{Interaction between ice sheets, ocean, and atmosphere on the Antarctic and Greenland ice sheets. \cite{b1}\\}
    \label{fig:background_interaction}
\end{figure*}

Predictions from climate models are improved when obtained via multi-model ensembles. The result is more accurate and produces more reliable predictions with lower uncertainty. The earliest works generally used an average of the simulated predictions to obtain the final result. However, this method would not entirely capture the prediction skill of the models in different regions. More recent approaches \cite{b7, b8} use a global weighting scheme based on regression or deep learning methods to reduce the discrepancy between observations and the weighted prediction obtained from the simulations. These models assign greater weight to more reliable predictions on a global scale but do not adequately handle the spatial variability arising from regional factors which affect changes in sea level. They can approximate the thermal expansion of the oceans but cannot properly account for regional factors like tsunamis, tides, boundary waves, river runoff, vertical land movement, etc. The spatial variability arising from such regional factors is a significant challenge that can lead to different weighting schemes based on the model's reliability in that specific region.

This paper explores a zonal regression model to address spatial variability in sea-level rise prediction. This approach assigns weights to models based on their prediction skills in a specific region, looking for improved predictions. Traditional weighting strategies use an inverse of the root mean square between the observations and the simulations to assign the weights on a global scale. This paper proposes a spatial-variability-aware regression model to learn the weights from historical predictions in a region. Empirical results in Figure \ref{fig:rmse} using root mean squared errors from the different model predictions, the baseline method, and the proposed approach display the effectiveness of the approach in reducing the prediction error.

\section{Background}
\label{section:background}

Global sea levels can be affected by factors like warming oceans and the addition of freshwater from continental ice (Figure \ref{fig:background_interaction}). Ocean dynamics can bring warm waters in contact with glaciers, leading to the decay of ice sheets. Changes in atmospheric temperature can increase surface temperatures over continents leading to further loss of ice sheets. This reduction in the ice mass balance can impact the sea level in different regions of the earth. Other factors such as tsunamis, tides, boundary waves, river runoff, and vertical land movement can also impact regional sea levels.

\begin{figure*}[ht]
    \centering
    \includegraphics[width=0.9\textwidth]{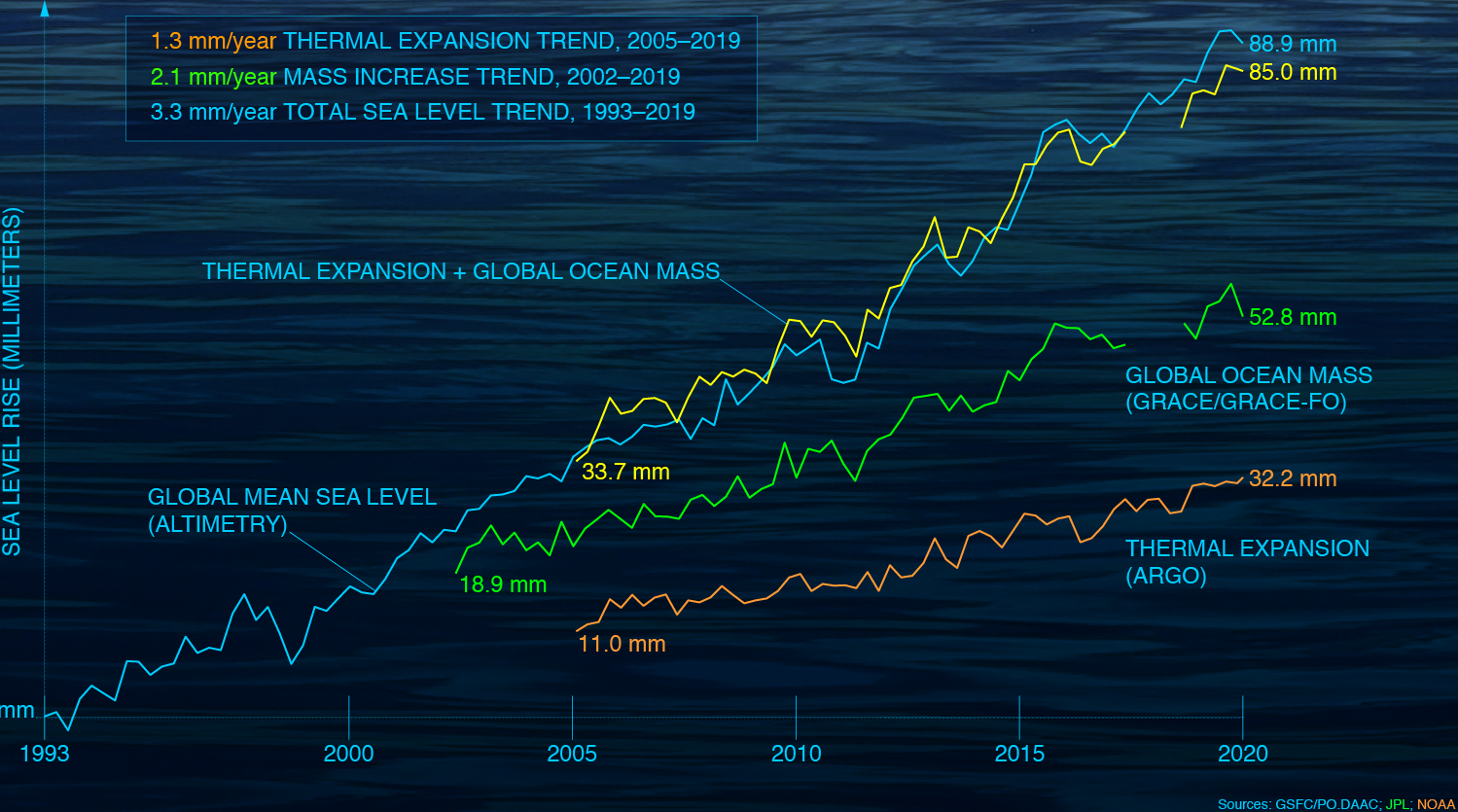}
    \caption{Contribution to sea-level rise from thermal expansion and ocean mass. \cite{b6}}
    \label{fig:Nasa_proj}
\end{figure*}

Figure \ref{fig:Nasa_proj} provides a comparison of the significant factors contributing to changes in the global mean sea level (GMSL). Different satellite systems measure different GMSL properties. For example, mass-driven changes such as ice melt and land water storage are measured by satellite altimeters, like GRACE/GRACE-FO \cite{b15}. Argos \cite{b16} measures ocean heat content by monitoring factors such as temperature and salinity changes in the ocean.

\section{Proposed Approach}
\label{section:methods}
This paper explores a zonal regression method to represent spatial variability in weighting multi-model ensemble climate projections. CMIP-6 projections obtained from the National Center for Atmospheric Research (NCAR) are used as the simulated data, while Copernicus sea-level anomaly gridded data are used for the ground truth observation.

Given a set of spatio-temporal tensors representing climate model projections ($P_r$) over a region $r$, the proposed weighting strategy aims to find a set of coefficients ($w_r$) for the multi-model ensemble such that the combination of projections with weights ($\sum_{i}P_{r,i}w_{r,i}$) improves the accuracy of the sea-level rise prediction while simultaneously reducing the prediction uncertainty. Here, $i$ represents the model under consideration.

We formulate the problem as a non-negative least squares problem where the goal is to find the set of weights $w_r$ that minimizes the prediction error:
\begin{align}
\operatorname*{argmin}_{w_{r,i}} ||O_{r} - \sum_{i}P_{r,i}w_{r,i}||,
\end{align} 
where $w_{r,i}>=0$ and $\sum_{i}w_{r,i}=1$. $O_r$ represent the ground truth observation data over region $r$.

\begin{figure}[ht]
    \centering
    \includegraphics[width=\linewidth]{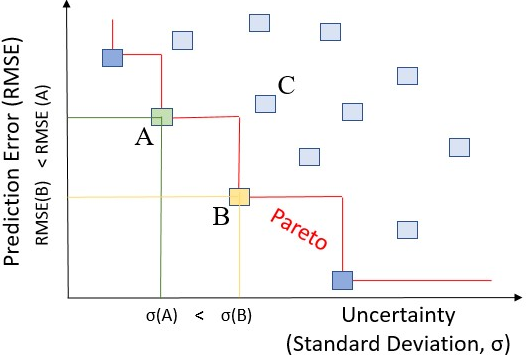}
    \caption{Pareto frontier for simultaneous reduction of prediction error and uncertainty.\\}
    \label{fig:pareto}
\end{figure}

Simultaneously, we also need to reduce the prediction uncertainty:
\begin{align}
minimize \sqrt{\frac{1}{T-1} \sum_{t=1}^{T} (E_{t} - \bar{E})^2},
\end{align}
where $E_{t} = ||O_{r,t} - \sum_{i}P_{r,t,i}w_{r,i}||$. This can be considered as a multi-objective optimization problem in which we want to minimize both the prediction error and the prediction uncertainty. One technique for solving this problem is to use Pareto optimality. The main goal in this technique is to find solutions such that each quantity (prediction error and prediction uncertainty) is minimized without affecting the other. Thus we want to find the Pareto frontier (i.e. set of Pareto efficient solutions) as in Figure \ref{fig:pareto}. A point $y'' \in \mathbb{R}^{m}$ is preferred to (dominates) another point $y' \in \mathbb{R}^{m}$ is referred to as $y'' \succ y'$. In Figure \ref{fig:pareto}, both $A$ and $B$ dominate $C$ in reducing the prediction error and the uncertainty.

Model similarity can lead to projections biased towards the largest set of similar models and the underestimation of uncertainties \cite{b23}. To address this, the proposed method down-weights models with a higher covariance in their predictions. The final weights obtained are a product of the weights obtained from the regression coefficients and the coefficients if the model is down-weighted. This zonal regression formulation overcomes the drawbacks of one-size-fits-all (OSFA) models such as GCMs which do not account for spatial variability.

\section{Experimental framework}

\subsection{Experimental Evaluation}
\label{section:experimental_evaluation}
\begin{figure*}[ht]
    \centering
    \includegraphics[width=0.9\textwidth]{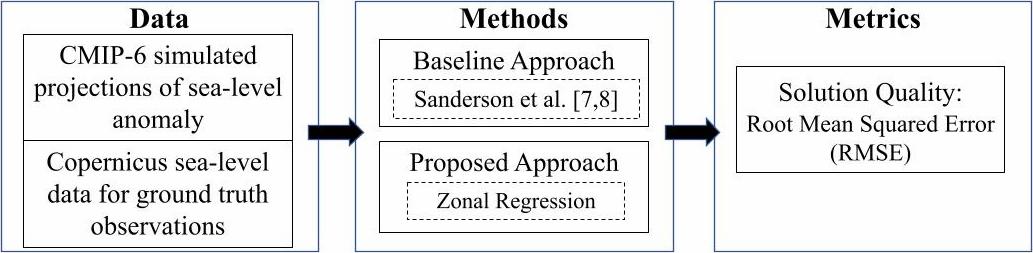}
    \caption{Experiment Design}
    \label{fig:ExperimentDesign}
\end{figure*}

A zonal regression model is employed to learn a weighting scheme for the simulated projections such that the output of the weighted predictions reduces the discrepancy with the observations. The main experimental goal was to compare the sea-level rise predictions obtained with the proposed spatial variability-aware zonal regression approach against the sea-level rise predictions from a one-size-fits-all method that doesn't adapt to zones. Figure \ref{fig:ExperimentDesign} shows the overall validation framework. The metric for comparison with the baseline \cite{b7, b8} was solution quality, specifically the root mean squared error between the combined prediction and the ground truth observations. The experiments were performed with the CMIP-6 projection models of sea-level height and the Copernicus satellite data for ground truth.

\vspace{5pt}

\vspace{-5pt}

\subsection{Data}
\label{section:datasets}
This work primarily uses Copernicus \cite{b9} sea-level gridded satellite data for ground truth observations and zos (Sea Surface Height Above Geoid) data from CMIP-6 (Coupled Model Intercomparison Project 6) for the simulated projections available from NCAR (National Center for Atmospheric Research).

The sea level satellite observation data \cite{b9} from Copernicus provides daily and monthly global estimates of mean sea level anomaly based on satellite altimetry measurements from 1993 to 2012. Here sea level anomaly refers to the height of the sea surface in comparison to the mean sea level at a given time and region. The dataset is generated using a two-satellite merged constellation where one satellite is used as a reference while the other is used to improve accuracy. 

The Coupled Model Intercomparison Project (CMIP) \cite{b10} is a collaborative framework for collecting, organizing, and distributing output from multiple coupled climate models performing common sets of experiments. It is focused on global climate model (GCM) simulations of the past, current and future climate. The major components include (1) DECK (Diagnostic, Evaluation, and Characterization of Klima), which are some common experiments, and CMIP historical simulations (1850–present); (2) common standards, coordination, infrastructure, and documentation; (3) ensembles of CMIP-Endorsed Model Intercomparison Projects (MIPs) specific to a phase of CMIP (CMIP6) that build upon the DECK and CMIP historical simulations. Some data variables of interest available from CMIP-6 include air pressure at sea level, air temperature, surface temperature, sea-level anomaly, surface runoff flux, etc. The experiments mentioned in this paper focus on sea-level anomaly (zos) data.

Experiments mentioned in this paper were focused on the Eastern North America (ENA) region due to the availability of more reliable ground truth data in the ENA region. The experiments were performed on four CMIP-6 models obtained from NCAR, namely CESM2 \cite{b17}, CESM2-FV2 \cite{b18}, CESM2-WACCM \cite{b19} and CESM2-WACCM-FV2 \cite{b20}.

\section{Preliminary results}
\label{section:results}
\begin{figure*}[ht]
    \centering
    \includegraphics[width=0.8\textwidth]{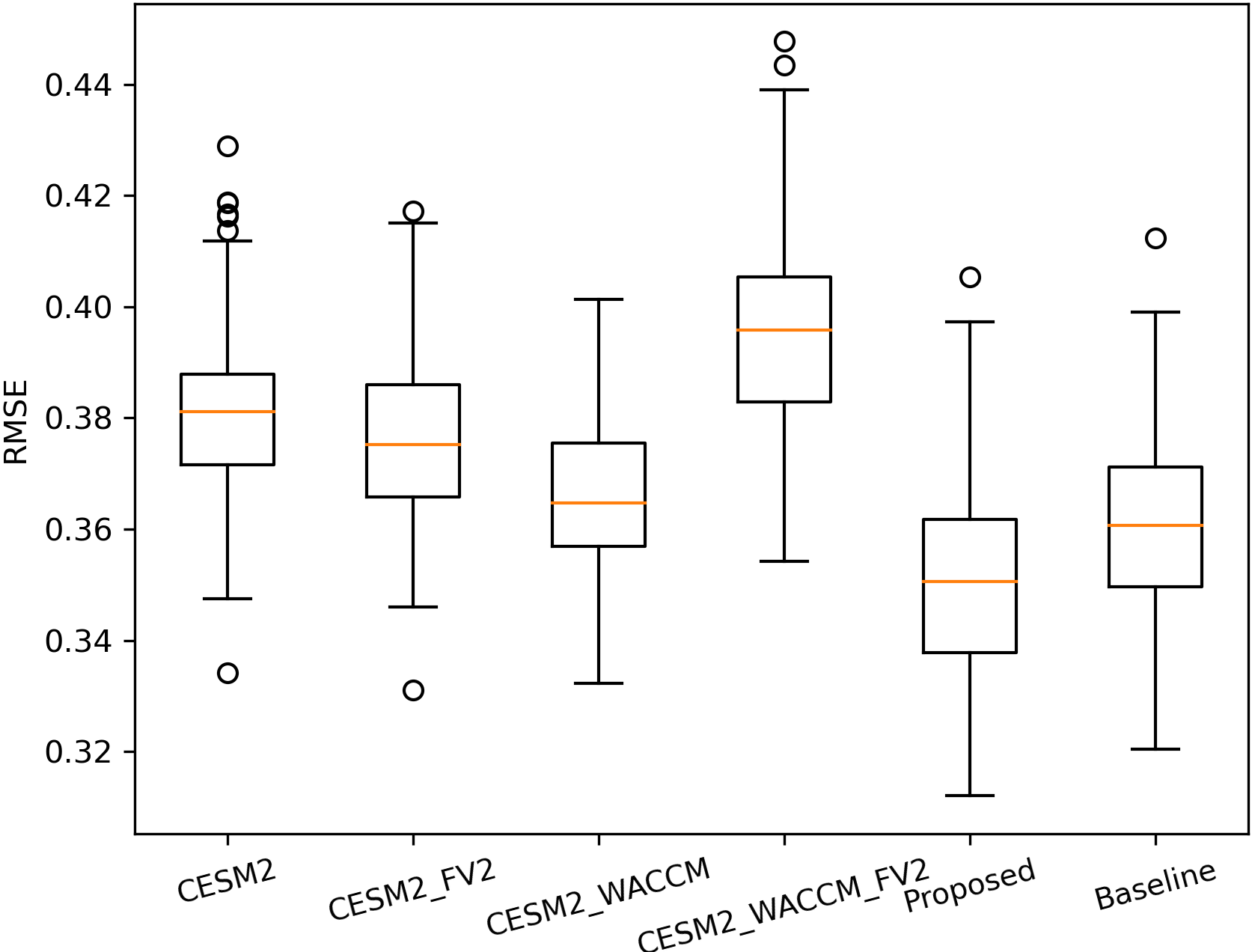}
    \caption{Root mean squared errors of each model.\\}
    \label{fig:rmse}
\end{figure*}

Figure \ref{fig:rmse} provides a comparison of the root mean squared errors (RMSE) in sea-level rise predictions from four simulation models, the baseline method, and the proposed approach. The RMSE obtained from each model prediction was higher than the weighted combination of the predictions. In the ENA region, the proposed method had a median RMSE of $0.35$ while the baseline method \cite{b7, b8} had a median RMSE of $0.36$ across the different time scales. The standard deviation of the prediction was $0.0177$ for the proposed method while the standard deviation of the baseline method was $0.0158$. The outliers in a boxplot represent datapoints that are located outside the whiskers of the box plot. For example, outliers in Figure \ref{fig:rmse} are data points $1.5$ times outside the interquartile range ($IQR$) above the upper quartile ($Q3$) and below the lower quartile ($Q1$), i.e. ($Q1 - 1.5 * IQR$ or $Q3 + 1.5 * IQR$). This shows that accounting for spatial variability leads to more reliable predictions.

\section{Related work and Discussion}
\label{related_work}
The Fourth National Climate Assessment (2018) mentions a weighting strategy used by the Intergovernmental Panel on Climate Change for differential weighting of climate projection models. The approach, based on works by Sanderson et al. \cite{b7, b8}, incorporates both skills in climatological performance and the inter-dependency of models due to factors such as common parameterizations. A set of weights are obtained via a ranking scheme of the climate models based on the inverse of the root mean square error (RMSE) between the observations and the predictions. The final output is a single set of weights that can be used to find a weighted average of the climate projections.
Inter-dependency between models is determined based on inter-model RMSE between each pair of models. A model is down-weighted if its inter-model RMSE with another model is significantly lower than a predetermined threshold. The final weight for each model is a product of its skill and its inter-dependency weights. This final set of weights obtained is used for combining projections irrespective of region and thus the strategy ignores the spatial variability from various regional physical forcings.

Spatial variability is receiving increasing attention in pattern detection \cite{b12, b13, b27, b28} and prediction  studies \cite{b11, b29}. Recent approaches \cite{b14} towards predicting regional sea-level rise with machine learning require explicit input values for the regional physical factors, which might be difficult to obtain in certain cases. Such bottlenecks can be avoided by employing the proposed approach.

\section{Conclusion and Future Work}
\label{section:conclusion_future_work}
This paper discussed a new spatial variability-aware zonal regression approach towards weighting sea-level rise models. Rather than relying on a model's global projection skill, this method learns a weighting scheme that accounts for regional factors in sea-level rise predictions.

\textbf{Future Work:} 
We plan to extend this work into more sophisticated approaches such as physics informed models using GWR (geographically weighted regression) or spatial-variability-aware deep learning methods \cite{b11} and work on novel ways to handle the inter-dependency between the models. A potential future direction of this work would be to make sea-level rise predictions at a higher resolution.

Sea-level rise is directly tied to the rise in greenhouse gas emissions. Therefore, we can improve the quantification of uncertainty in sea-level rise projections by considering different scenarios that are likely to affect levels of greenhouse gas emissions in the future. For example, with no human intervention, scientists project emissions will double by 2100. If all countries comply with the Paris Agreement, emissions should reach net zero after 2050. A third scenario assumes countries only meet their current national commitments for reducing emissions. In our future work, we plan to explore how such emissions scenarios can be used to reduce uncertainty in sea-level rise projections.

\section*{Acknowledgment}

This material is based upon work supported by the National Science Foundation under Grants No. 2118285, 2040459, 1901099, and 1916518. We also thank  Kim Koffolt, the iHarp community, and the Spatial Computing Research Group for valuable comments and refinements.


\end{document}